\documentclass[twocolumn,prb,showpacs,amsmath,amssymb,superscriptaddress]{revtex4}

\pdfoutput=1

\usepackage{bm}
\usepackage{graphicx}
\usepackage{amssymb}
\usepackage{xcolor}
\newcommand{\beg}{\begin{equation}}
\newcommand{\en}{\end{equation}}

\begin{document}

\title{Heat capacity of URu$_{2-x}$Os$_x$Si$_2$ at low temperatures}

\author{D. L. Kunwar }  
\affiliation{Department of Physics, Kent State University, Kent, OH 44242, USA}

\author{S. R. Panday}  
\affiliation{Department of Physics, Kent State University, Kent, OH 44242, USA}

\author{Y. Deng}
\affiliation{Center for Advanced Nanoscience, University of California, San Diego, La Jolla, CA 92093, USA}
\affiliation{Department of Physics, University of California at San Diego, La Jolla, CA 92903, USA}

\author{S. Ran} 
\affiliation{Department of Physics, Washington University in St. Louis, St. Louis, MO 63130, USA}

\author{R. E. Baumbach} 
\affiliation{National High Magnetic Field Laboratory, Florida State University, Tallahassee, FL 32310; Department of Physics, Florida State University, Tallahassee, FL 32306, USA}

\author{M. B. Maple}
\affiliation{Center for Advanced Nanoscience, University of California, San Diego, La Jolla, CA 92093, USA}
\affiliation{Department of Physics, University of California at San Diego, La Jolla, CA 92903, USA}

\author{Carmen C. Almasan}
\affiliation{Department of Physics, Kent State University, Kent, OH 44242, USA}

\author{M. Dzero}
\affiliation{Department of Physics, Kent State University, Kent, OH 44242, USA}

\begin{abstract}
We perform measurements of the heat capacity as a function of temperature on URu$_{2-x}$Os$_x$Si$_2$ alloys.  Our experimental results show that the critical temperature of the second-order phase transition increases while the value of the Sommerfeld coefficient in the ordered state decreases with 
an increase in osmium concentration. We also observe the increase in the values of the heat capacity at the critical temperature as well as a broadening
of the critical fluctuations region with an increase in $x$. We analyze the experimental data using the Haule-Kotliar model which, in particular, identifies the 
'hidden order' transition in the parent material URu$_2$Si$_2$ as a transition to a state with nonzero hexadecapole moment. We demonstrate that our experimental 
results are consistent with the predictions of that model.
\end{abstract}


\date{\today}

\maketitle

\section{Introduction}  
The manifestly second-order phase transition develops in URu$_2$Si$_2$ at a temperature $T_{\textrm{c0}}=17.55$ K.\cite{HOMydosh1985,EarlyExp1986,EarlyExp1986Z,EarlyExp1987} The nature of the order parameter emerging below the transition has remained hotly debated during the last 25 years.\cite{EarlyTheory1,EarlyTheory2,EarlyTheory3,EarlyTheory4,Review2014,Review2020} In particular, the problem with the identification of the order parameter has made an analysis of  thermodynamic data quite challenging. Notably, the temperature dependence of the heat capacity, for example, shows that the Ginzburg region, where the contribution from critical fluctuations is comparable or exceeds the mean-field contribution, is very narrow and yet the absence of a clear idea about the nature of the order parameter inhibits any attempt to analyze the thermodynamic data even at the mean-field level. 

Despite the fact that there was a significant progress made both experimentally and theoretically towards the identification of the 'hidden order',\cite{Review2014,Review2020} new ideas about the nature of the 'hidden order' transition still continue to appear.\cite{KH2009,Recent1,Recent2,Recent3,Recent4,Recent5,Recent6,Hastatic2013} Specifically, one of the main focus points of the ongoing discussions is still  whether the 'hidden order' transition involves quasi-localized $5f^2$ states of uranium or it is driven by the itinerant electronic degrees of freedom, which are hybridized with the $5f$ electronic states. The possible resolution of this debate is likely to come from designing the experiments in a way that would help one to unambiguously contrast the measurements results with theoretical predictions.\cite{Maple2010,Fano2012,Lu2012,Girsh2015}

\begin{figure}[ht]
\centering
\includegraphics[width=0.85\linewidth]{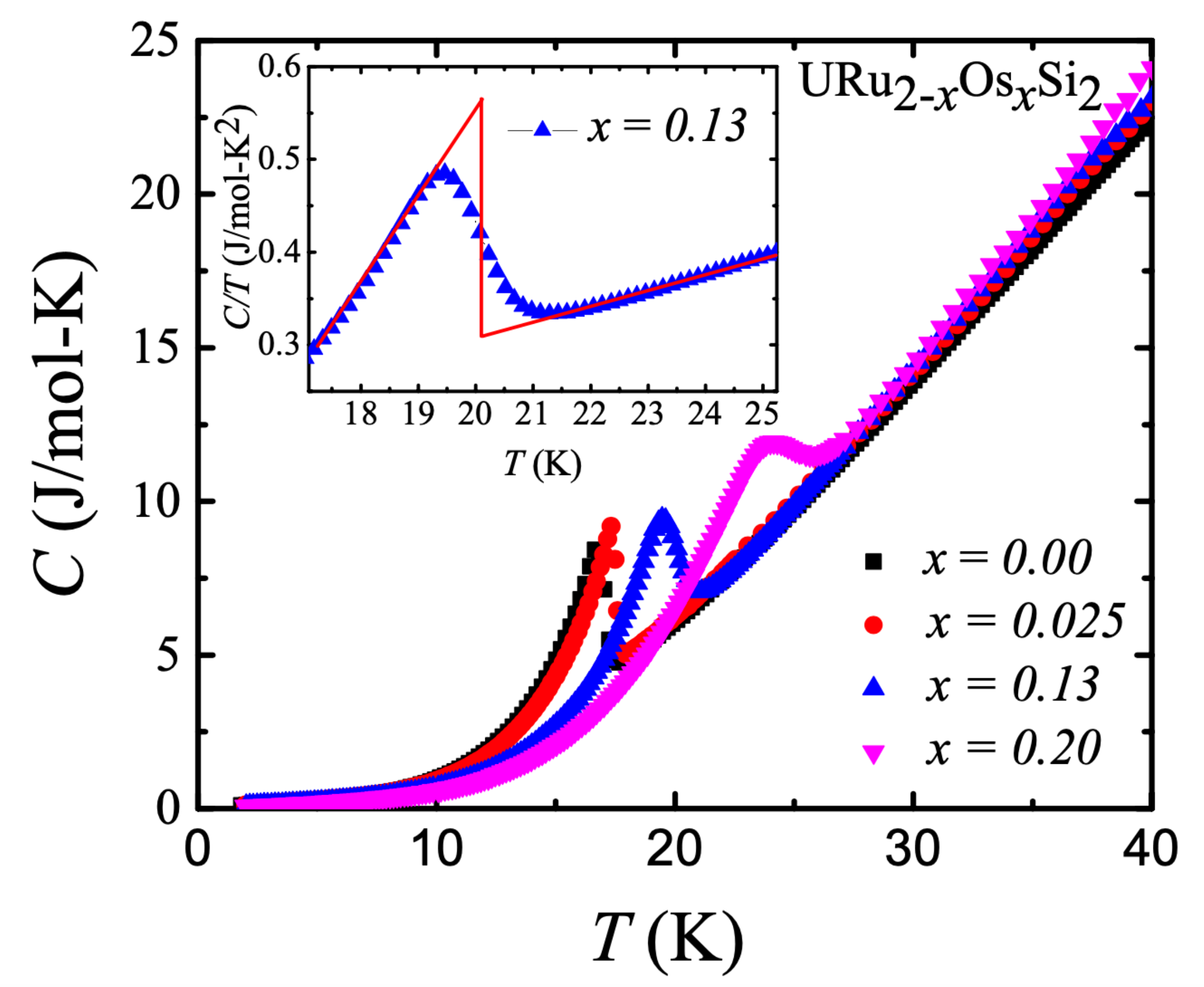}
\caption{Heat capacity $C$ as a function of temperature $T$ in URu$_{2-x}$Os$_x$Si$_2$ alloys. Inset: Entropy construction to estimate the critical temperature of the second-order phase transition. There are three main features that are of interest to us: the critical temperature $T_c$, the maximum value of the heat capacity, and the broadening of the thermal fluctuations region around $T_c$ are all increasing with increasing osmium concentration $x$.}
\label{Fig1}
\end{figure}

One type of such experiments involves the studies of alloys in which ruthenium (Ru) is substituted either with iron (Fe) or osmium (Os).\cite{PNAS1,PNAS2,PNAS3} For example, the substitution of Ru with Os produces an effect of negative pressure leading to a predominantly 
isotropic lattice expansion. By contrasting the changes in the critical temperature $T_c$ with the changes in the Sommerfeld coefficient $\gamma$, one should (at least, in principle) be able to check whether this leads to a contradiction with either 'itinerant' or 'localized' theoretical models.
Qualitatively, in the itinerant scenario, for example, one would expect that both $\gamma$ and $T_c$ should either decrease or increase with the change in external pressure, since both of these quantities provide a measure of hybridization between the itinerant and localized degrees of freedom. 

In this Letter, we report on the measurements of the heat capacity in URu$_{2-x}$Os$_x$Si$_2$ across the 'hidden order' phase transition. Our experimental results demonstrate that the critical temperature of the transition grows with an increase in osmium concentrations and reaches the value of $T_c\approx 1.43T_{\textrm{c0}}$ for $x=0.2$. Consequently, we have also analyzed the electronic contribution to the heat capacity at low enough temperatures below $T_c$ and found that the Sommerfeld coefficient actually decreases with $x$. We were also able to obtain satisfactory fits for the heat capacity data using the model put forward by Haule and Kotliar,\cite{KH2009,KH2010} which explains the second order transition as a state with a complex order parameter: the real part of the order parameter corresponds to the nonzero hexadecapole moment, while the imaginary part is determined by the staggered magnetic moment. Taking into account the results of the earlier transport measurements in URu$_{2-x}$Os$_x$Si$_2$, we arrived to the conclusion that osmium substitutions promote the emergence of the antiferromagnetic order at concentrations $x\geq0.15$.

\section{Experimental details} 
Single crystals of URu$_{2-x}$Os$_x$Si$_2$ were grown by the Czochralski method in a tetra-arc furnace. 
The crystals were cut into a rectangle with the $c$-axis  along the shortest dimension of the crystal. These single crystals were first polished with sandpaper to make their surface smooth for better coupling to the heat capacity platform.  They were then washed thoroughly with ethanol to remove any impurities left.

Heat capacity measurements were performed in zero magnetic field over the temperature range $T=2\div50$ K and the data were obtained using a relaxation technique in the He-4 option of the Quantum Design Physical Properties Measurement System.  
\section{Data analysis} 
The analysis of the heat capacity data at temperatures much lower than the critical temperature, $T\ll T_c$, is straightforward since the main contribution to the heat capacity 
in this temperature region comes from itinerant electrons and the lattice vibrations, $C_{\textrm{el}}(T)$+$C_{\textrm{ph}}(T)$. We find that at low temperatures 
$C_{\textrm{el}}(T)\approx\gamma T$ and $C_{\textrm{ph}}(T)=(T/\omega_D)^3{\cal I}(\omega_D/T)$, where $\gamma$ is the Sommerfeld coefficient, $\omega_D$ is the Debye temperature,
and ${\cal I}(x)$ is a known function of $x$: ${\cal I}(x\to\infty)\approx 26$. 

At temperatures $T\sim T_c$, however, the contribution from the electronic degrees of freedom, which govern the 'hidden order' transition, becomes a dominant one. In order to analyze our data in this temperature region, one generally needs to put forward an idea about the origin of the order parameter so that the temperature dependence of the heat capacity can be computed and ultimately be compared to (or used to fit) the experimental data. Thus, as a starting point, we need to decide on whether to consider the itinerant degrees of freedom (i.e. electrons on the $spd$-orbitals of uranium) as a driving force for the second-order phase transition or, on the contrary, adopt the 'localized picture' in which the electrons on the localized $5f^2$ orbitals are the main driving force for the transition. Based on the earlier observations of BCS-like features of the transition in the stoichiometric URu$_2$Si$_2$ \cite{HOMydosh1985} it is, indeed, tempting to use one of the recently proposed itinerant models (for a recent review see [\onlinecite{Review2014}] and references there in) to analyze the data. One needs to keep in mind, however, that the mean-field-like temperature dependence of the thermodynamic response functions in the BCS model \cite{BCS} (excluding the very narrow region in the immediate vicinity of the transition) is controlled by the retarded nature of the electron-phonon interactions that drive the superconducting transition.\cite{Eliashberg1986} This is manifested in the smallness of the ratio of the Debye frequency to the Fermi energy, $\omega_D/\varepsilon_F\ll 1$. \cite{Kos2004,Joerg1} In particular, in the weak coupling limit, the retardation effects are implemented as an ultraviolet cutoff 
$\Lambda=\omega_D$ in the self-consistent calculation of the order parameter. There are known limitations to this line of arguments, \cite{Joerg2} but we think that these limitations are not relevant for the 'hidden-order' transition. To summarize, to the best of our knowledge there is no experimental evidence for the retardation effects being observed through the 'hidden order' transition and the emergence of an energy scale (analogous to $\omega_D$ in the BCS theory) that would be much smaller than the Fermi energy.

Therefore, we are compelled to adopt the point of view in which interactions between the localized $f$-orbital degrees of freedom must be the ones leading to the transition.\cite{Shen2018} Among the multiple choices of models available to us (see e.g. \cite{Review2014,Review2020} for review) to analyze our data we consider the model proposed by Haule and Kotliar (HK). \cite{KH2009,KH2010} The HK model involves a system of interacting two-level systems (TLSs) and each TLS corresponds to a ground-state non-Kramers doublet of uranium $5f^2$ valence configuration. The components of the corresponding state vector $\vert \Psi\rangle^T=(\vert a\rangle ~\vert b\rangle)$ of the non-Kramers doublet are 
\beg\label{Eq1}\nonumber
\begin{split}
\vert a\rangle&=\frac{i}{\sqrt{2}}(\vert 4\rangle-\vert -4\rangle), \\
\vert b\rangle&=\frac{\cos\phi}{\sqrt{2}}(\vert 4\rangle+\vert -4\rangle)-\sin\phi\vert 0\rangle.
\end{split}
\en
Here the states are written in the eigenbasis of the total angular momentum operator, $\hat{J}^2$, and its projection on the $z$-axis, $\hat{J}_z$
(we remind the reader that the $5f^2$ valence configuration corresponds to a state with $J=4$). 
One can check that the following averages acquire non-zero values: $\langle b\vert\hat{J}_z\vert a\rangle=4i\cos\phi$ and $\langle b\vert(\hat{J}_x\hat{J}_y+\hat{J}_y\hat{J}_x)(\hat{J}_x^2-\hat{J}_y^2)\vert a\rangle=12\sqrt{35}\sin\phi+28\cos\phi$. Here $\phi$ is a parameter that ultimately determines the magnitude of the magnetic moment at  $T\ll T_c(x=0)$ in the 'hidden order' state. Thus, it is convenient to formally associate the first average with $\langle\Psi\vert\hat{\sigma}_y\vert\Psi\rangle$ and the second average with $\langle\Psi\vert\hat{\sigma}_x\vert\Psi\rangle$, where $\hat{\sigma}_x$ and $\hat{\sigma}_y$ are the Pauli matrices. As a result, one can formally introduce the complex order parameter $\psi=\langle\Psi\vert\hat{\sigma}_x+i\hat{\sigma}_y\vert\Psi\rangle$ describing a state that may emerge below some temperature as the lowest energy state for the following Hamiltonian:
\beg\label{Eq2}
\begin{split}
\hat{H}&=-\sum\limits_{i\not =j}V({\vec r}_i-{\vec r}_j)\hat{\sigma}_x({\vec r}_i)\hat{\sigma}_x({\vec r}_j)-\frac{\Delta}{2}\sum\limits_i
\hat{\sigma}_z({\vec r}_i)\\&+\sum\limits_{ij}U({\vec r}_i-{\vec r}_j)\hat{\sigma}_y({\vec r}_i)\hat{\sigma}_y({\vec r}_j).
\end{split}
\en
Here the first term describes the interactions which lead to the emergence of the 'hidden order' ($V>0$, a state with non-zero hexadecapole moment), the second term accounts for the splitting of the non-Kramers doublet, while the third term accounts for the antiferromagnetic interactions ($U>0$). 
As it has been recently shown,\cite{KH2010} the mean-field analysis of the model Hamiltonian (\ref{Eq2}) provides a satisfactory description of the thermodynamics measurements under external pressure and magnetic field. 
\begin{figure}[h]
\centering
\includegraphics[width=1.0\linewidth]{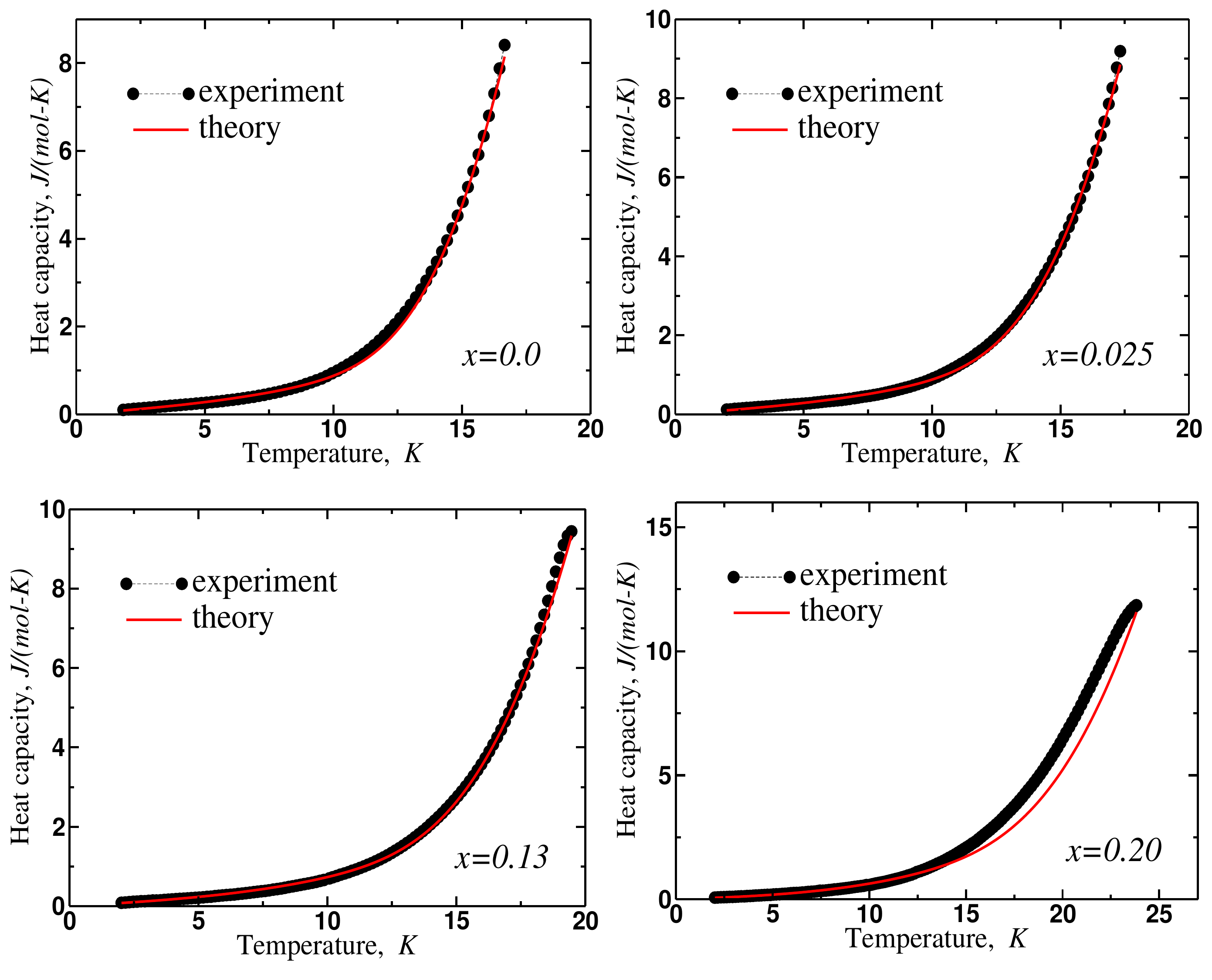}
\caption{Results of our analysis of the heat capacity at temperatures below the critical temperature. To fit the data on the panel for $x=0$, we used the following values of the parameters: $n_{\textrm{ph}}=375$, $n_{\textrm{tls}}=9.15$, $\gamma=0.049$ (J/mol$\cdot${K}$^2$), 
$\omega_D=305$ K, $V_0=18.25$ K and $s(T=0)=3.85$.
To fit the remaining data, we have only changed the values of $\gamma$ and $V_0$. Panel for $x=0.025$: $\gamma=0.049$ (J/mol$\cdot${K}$^2$), 
$V_0=18.75$. Panel for $x=0.13$: $\gamma=0.038$ (J/mol$\cdot${K}$^2$), 
$V_0=21.15$. Panel for $x=0.20$: $\gamma=0.029$ (J/mol$\cdot${K}$^2$), 
$V_0=25.75$ K.}
\label{FigDataFits}
\end{figure}

To perform the data analysis, we introduce several simplifications of the model Hamiltonian (\ref{Eq2}). First and foremost we assume that the interactions leading to the emergence of the nonzero expectation value $\langle\Psi\vert\hat{\sigma}_x\vert\Psi\rangle$ are sufficiently long-ranged with the characteristic length scale $r_0\gg a$ ($a$ is the lattice spacing). This allows us to perform a controlled calculation of the heat capacity, so that the contribution from thermal fluctuation effects are small in powers of $(a/r_0)^3$. \cite{VLP} Secondly, we completely neglect the contribution from the last two terms: the second term is practically irrelevant for the temperatures $T\sim T_c$, while, as we will see from our analysis below, the third term contributes significantly to the heat capacity only at relatively high osmium concentrations, $x\sim 15\%$.

In order to compute the specific heat, we follow the avenue of Ref. \onlinecite{VLP}. We introduce $s=\langle\Psi\vert\hat{\sigma}_x({\vec r}_i)\vert\Psi\rangle$ and, on the account of our discussion above, we re-write (\ref{Eq2}) as follows:
\beg\label{Eq3}
\begin{split}
\hat{H}&=NV_0s^2-sV_0\sum\limits_{i}\hat{\sigma}_x({\vec r}_j)\\&-\sum\limits_{i\not =j}V({\vec r}_i-{\vec r}_j)(\hat{\sigma}_x({\vec r}_i)-s)(\hat{\sigma}_x({\vec r}_j)-s),
\end{split}
\en
where $V_0=\sum_jV({\vec r}_i-{\vec r}_j)$ and $N$ is the number of uranium lattice sites. Within the mean-field approximation the third term in (\ref{Eq3}) can be neglected as it describes the effect of thermal fluctuations to the mean-field results. The subsequent minimization of the free energy with respect to the mean field parameter $s$ ultimately results in the following expression for the heat-capacity ($k_B=1$, $\beta=1/T$)
\beg\label{Eq4}
c_{\textrm{h.o.}}(T)=\frac{\left(\beta V_0s(T)\right)^2}{\cosh^2[\beta V_0s(T)]-\beta V_0}
\en
and the temperature dependence of the order parameter $s(T)$ can be found by solving the mean-field equation $s=\tanh(\beta V_0s)$.

Thus, we write $C(T)=C_{\textrm{el}}(T)+C_{\textrm{ph}}(T)+C_{\textrm{h.o.}}(T)$ and employ this expression to fit our heat-capacity data. Specifically, we use the data for the stoichiometric compound to find the corresponding pre-factors for the phonon $C_{\textrm{ph}}(T)=n_{\textrm{ph}}c_{\textrm{ph}}(T)$ and $f$-electron $C_{\textrm{h.o.}}(T)=n_{\textrm{tls}}c_{\textrm{h.o.}}(T)$ contributions along with the Debye temperature $\omega_D$ and dimensionless matrix element $s(T=0)$. 

The results of our analysis are shown in Fig. \ref{FigDataFits}. Generally, we find that the heat capacity data for $x=0$, $x=0.025$, and $x=0.13$ can be systematically described by our model (\ref{Eq3}). After we found a satisfactory result by fixing the lattice parameters obtained from the analysis of the heat capacity data at $T\leq T_{c}$ for URu$_2$Si$_2$, we analyze the remaining sets by changing only two parameters: interaction strength $V_0$ and Sommerfeld coefficient $\gamma$. 

\begin{figure}[ht]
\centering
\includegraphics[width=0.85\linewidth]{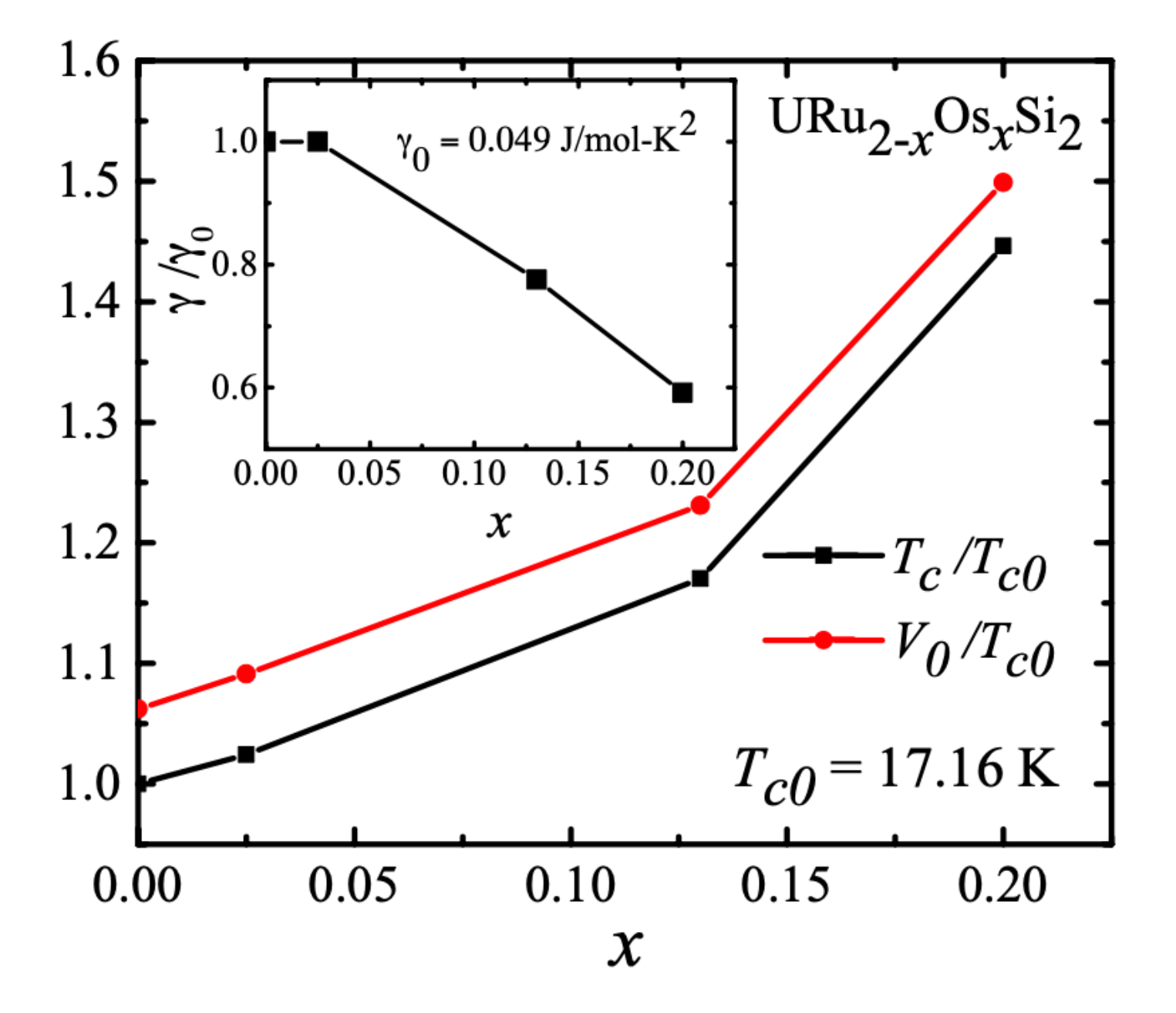}
\caption{Main panel: Dependence of the interaction strength $V_0$ and the critical temperature of the second phase transition as a function of osmium concentration $x$. Given the overall mean-field nature of the transition, it is not surprising to find $V_0\propto T_c$. Inset: Extracted value of the Sommerfeld coefficient $\gamma$ as a function of the osmium concetration $x$. The reduction in the value of $\gamma$ signals the decrease in the hybridization between the $spd$- and $f$-electron states of uranium.}
\label{Fig3}
\end{figure}

Since we have observed experimentally that the critical temperature of the transition increases with $x$ and also from the fact that at the level
of the mean-field approximation $T_c\simeq V_0$ it follows that in order to obtain good fits to the data we need to increase the value of $V_0$.
At the same time we have also noted that the best fits are obtained if we gradually decrease the value of $\gamma$ with increasing $x$. These results 
are summarized in Fig. \ref{Fig3}

Our fitting procedure did not work well for the $x=0.2$ sample: there are obvious descripancies between the theoretically predicted result and the
experimental data in the temperature region where the contribution from the $f$-electrons becomes comparable with the remaining contributions from the 
itinerant electrons and the lattice vibrations. Our interpretation of these observations is discussed in what follows. 

\section{Discussion} 
\begin{figure}[ht]
\centering
\includegraphics[width=0.85\linewidth]{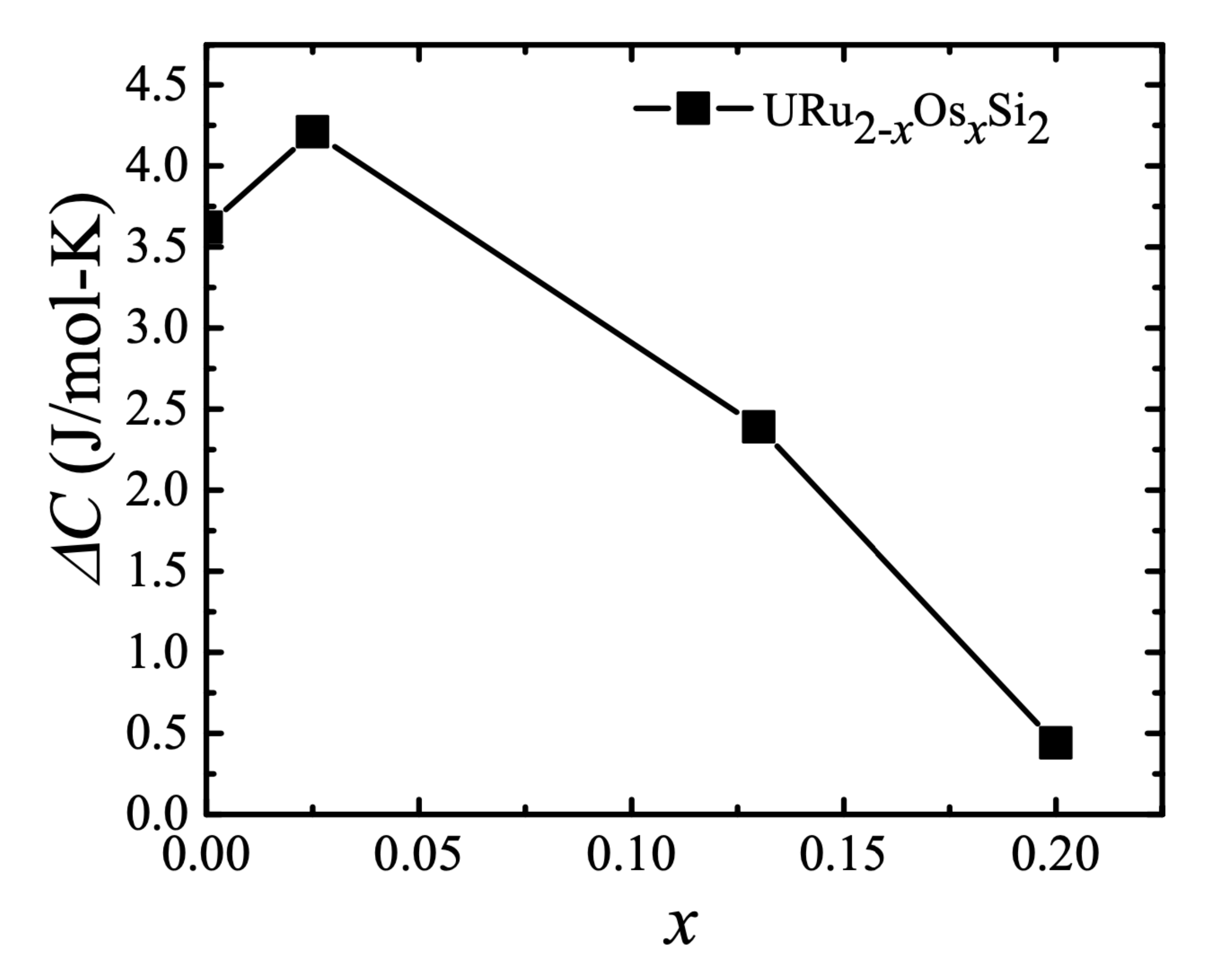}
\caption{Specific heat jump at the critical temperature, $\Delta C(T_c)=C_{\textrm{max}}(T_c-\epsilon)-C_{\textrm{min}}(T_c+\epsilon)$ ($\epsilon\ll T_c$), as a function of osmium concentration $x$.}
\label{Fig4Jump}
\end{figure}

The possible origin of the discrepancy between the results of our fitting procedure and the heat capacity for the $x=0.2$
can be traced back to the recent results of the transport studies of URu$_{2-x}$Os$_x$Si$_2$. \cite{PNAS1} Specifically, it was shown that at approximately $x\approx 0.15$ there is a change in slope of $T_c$ vs. $x$ corresponding to the onset of the long-range antiferromagnetic order. 
In addition, we note that the value of the heat capacity for $x=0.2$ at $T=T_c$ exceeds the corresponding values for other samples by approximately 
20\%. This fact can be accounted for by the enhanced exchange interactions $U({\vec r}_i-{\vec r}_j)$. In addition, our data clearly show that there is 
an increasing contribution $\propto (a/r_0)^3\sqrt{T_c/|T-T_c|}$ from the thermally induced magnetic fluctuations with the correlation function $D_{ij}=\langle(\hat{\sigma}_y({\vec r}_i)-\mu_s)(\hat{\sigma}_y({\vec r}_j)-\mu_s)\rangle$, where $\mu_s$ is the value of the staggered magnetic moment. These observations are supported by the decrease of the specific heat jump at $T=T_c$, Fig. \ref{Fig4Jump}, which is again accounted for by the broadening of the Ginzburg region, which we estimate to be $|\delta T|/T_c\simeq (a/r_0)^6$. In other words, the expansion of the lattice with $x$ must cause the reduction of $r_0$ and the broadening of the critical fluctuations region. 

Another consequence of our analysis is the observation of the decreasing value of the Sommerfeld coefficient $\gamma(x)$ with $x$, Fig. \ref{Fig3} (inset). This reduction 
most likely is induced by the diminished hybridization between the $spd$ and $5f^2$ states. Since the effect of osmium substitution produces the lattice expansion reduction in hybridization implies that the leading fluctuation channel is $5f^2\leftrightarrow 5f^1$, as it was suggested in Ref. \onlinecite{Hastatic2013}.  This behavior is in contrast with the option of valence fluctuations in the second channel $5f^2\leftrightarrow 5f^3$, which would actually be enhanced with increase in $x$ and, consequently, lead to the increase in the values of $\gamma$. 

Lastly, we would like to comment on the increasing value of $V_0$ which follows from our fitting procedure. We remind the reader that the parameter $V_0$ plays the role of the effective interaction strength and is equal to an integrated over space $V({\mathbf r})$, Eq. (\ref{Eq3}). Earlier studies \cite{HOMydosh1985,Pinaki2015,PNAS2,PNAS3} have associated the product $V_0s(0)$ with the hybridization (and/or charge) gap. In our theory $s(0)$ cannot possibly change with $x$ or external pressure since its value is determined by a local matrix element. However, since the lattice is expanding, $V_0$ must be changing and its apparent increase with $x$ implies that $V_0$ must be inverse-proportional to the correlation radius $r_0$. 


\section{Conclusions} We have measured and analyzed the heat capacity as a function of temperature in URu$_{2-x}$Os$_x$Si$_2$. Osmium alloying leads to a lattice expansion equivalent to an effect of negative pressure. Our data may be considered as an addition to the transport data on the same alloys, \cite{PNAS1} with similar conclusions regarding the increase in $T_c$ with $x$. Decreasing values of the Sommerfeld coefficient with the increase in $x$ implies that the second-order phase transition is driven by the interactions between the localized $f$-orbital degrees of freedom. 
The second novel aspect of our work consists in 
an attempt to use a theoretical model to analyze our results: this is in clear contrast with the earlier experimental studies which used purely phenomenological expressions \cite{HOMydosh1985,PNAS1,Pinaki2015,PNAS2,PNAS3} to estimate the changes in the heat capacity through the second order transition.

\section{Acknowledgments} The work at Kent State University was financially supported by the National Science Foundation under grants DMR-1904315 (D.L.K. and C.C.A.) and NSF-DMR-2002795 (S.R.P. and M.D.). Research at the University of California, San Diego was supported by the U.S. Department of Energy (DOE), Office of Science, Basic Energy Sciences (BES), under Grant No. DE-FG02-04ER46105 (single crystal growth), and by the National Science Foundation (NSF) under Grant No. DMR-1810310 (physical properties measurements). Research at the National High Magnetic Field Laboratory was supported by NSF Cooperative Agreement DMR-1157490, the State of Florida, and the DOE. R.E.B. was supported by the Center for Actinide Science and Technology (CAST), an Energy Frontier Research Center (EFRC) funded by the U.S. DOE, BES, under grant no. DE-SC0016568
\bibliography{hdosmium1}

\begin{thebibliography}{34}
\expandafter\ifx\csname natexlab\endcsname\relax\def\natexlab#1{#1}\fi
\expandafter\ifx\csname bibnamefont\endcsname\relax
  \def\bibnamefont#1{#1}\fi
\expandafter\ifx\csname bibfnamefont\endcsname\relax
  \def\bibfnamefont#1{#1}\fi
\expandafter\ifx\csname citenamefont\endcsname\relax
  \def\citenamefont#1{#1}\fi
\expandafter\ifx\csname url\endcsname\relax
  \def\url#1{\texttt{#1}}\fi
\expandafter\ifx\csname urlprefix\endcsname\relax\def\urlprefix{URL }\fi
\providecommand{\bibinfo}[2]{#2}
\providecommand{\eprint}[2][]{\url{#2}}

\bibitem[{\citenamefont{Palstra et~al.}(1985)\citenamefont{Palstra, Menovsky,
  Berg, Dirkmaat, Kes, Nieuwenhuys, and Mydosh}}]{HOMydosh1985}
\bibinfo{author}{\bibfnamefont{T.~T.~M.} \bibnamefont{Palstra}},
  \bibinfo{author}{\bibfnamefont{A.~A.} \bibnamefont{Menovsky}},
  \bibinfo{author}{\bibfnamefont{J.~v.~d.} \bibnamefont{Berg}},
  \bibinfo{author}{\bibfnamefont{A.~J.} \bibnamefont{Dirkmaat}},
  \bibinfo{author}{\bibfnamefont{P.~H.} \bibnamefont{Kes}},
  \bibinfo{author}{\bibfnamefont{G.~J.} \bibnamefont{Nieuwenhuys}},
  \bibnamefont{and} \bibinfo{author}{\bibfnamefont{J.~A.}
  \bibnamefont{Mydosh}}, \bibinfo{journal}{Phys. Rev. Lett.}
  \textbf{\bibinfo{volume}{55}}, \bibinfo{pages}{2727} (\bibinfo{year}{1985}).

\bibitem[{\citenamefont{Maple et~al.}(1986)\citenamefont{Maple, Chen,
  Dalichaouch, Kohara, Rossel, Torikachvili, McElfresh, and
  Thompson}}]{EarlyExp1986}
\bibinfo{author}{\bibfnamefont{M.~B.} \bibnamefont{Maple}},
  \bibinfo{author}{\bibfnamefont{J.~W.} \bibnamefont{Chen}},
  \bibinfo{author}{\bibfnamefont{Y.}~\bibnamefont{Dalichaouch}},
  \bibinfo{author}{\bibfnamefont{T.}~\bibnamefont{Kohara}},
  \bibinfo{author}{\bibfnamefont{C.}~\bibnamefont{Rossel}},
  \bibinfo{author}{\bibfnamefont{M.~S.} \bibnamefont{Torikachvili}},
  \bibinfo{author}{\bibfnamefont{M.~W.} \bibnamefont{McElfresh}},
  \bibnamefont{and} \bibinfo{author}{\bibfnamefont{J.~D.}
  \bibnamefont{Thompson}}, \bibinfo{journal}{Phys. Rev. Lett.}
  \textbf{\bibinfo{volume}{56}}, \bibinfo{pages}{185} (\bibinfo{year}{1986}),
  \urlprefix\url{https://link.aps.org/doi/10.1103/PhysRevLett.56.185}.

\bibitem[{\citenamefont{Schlabitz et~al.}(1986)\citenamefont{Schlabitz,
  Baumann, Pollit, Rauchschwalbe, Mayer, Ahlheim, and Bredl}}]{EarlyExp1986Z}
\bibinfo{author}{\bibfnamefont{W.}~\bibnamefont{Schlabitz}},
  \bibinfo{author}{\bibfnamefont{J.}~\bibnamefont{Baumann}},
  \bibinfo{author}{\bibfnamefont{B.}~\bibnamefont{Pollit}},
  \bibinfo{author}{\bibfnamefont{U.}~\bibnamefont{Rauchschwalbe}},
  \bibinfo{author}{\bibfnamefont{H.~M.} \bibnamefont{Mayer}},
  \bibinfo{author}{\bibfnamefont{U.}~\bibnamefont{Ahlheim}}, \bibnamefont{and}
  \bibinfo{author}{\bibfnamefont{C.~D.} \bibnamefont{Bredl}},
  \bibinfo{journal}{Zeitschrift f{\"u}r Physik B Condensed Matter}
  \textbf{\bibinfo{volume}{62}}, \bibinfo{pages}{171} (\bibinfo{year}{1986}),
  \urlprefix\url{https://doi.org/10.1007/BF01323427}.

\bibitem[{\citenamefont{Broholm et~al.}(1987)\citenamefont{Broholm, Kjems,
  Buyers, Matthews, Palstra, Menovsky, and Mydosh}}]{EarlyExp1987}
\bibinfo{author}{\bibfnamefont{C.}~\bibnamefont{Broholm}},
  \bibinfo{author}{\bibfnamefont{J.~K.} \bibnamefont{Kjems}},
  \bibinfo{author}{\bibfnamefont{W.~J.~L.} \bibnamefont{Buyers}},
  \bibinfo{author}{\bibfnamefont{P.}~\bibnamefont{Matthews}},
  \bibinfo{author}{\bibfnamefont{T.~T.~M.} \bibnamefont{Palstra}},
  \bibinfo{author}{\bibfnamefont{A.~A.} \bibnamefont{Menovsky}},
  \bibnamefont{and} \bibinfo{author}{\bibfnamefont{J.~A.}
  \bibnamefont{Mydosh}}, \bibinfo{journal}{Phys. Rev. Lett.}
  \textbf{\bibinfo{volume}{58}}, \bibinfo{pages}{1467} (\bibinfo{year}{1987}),
  \urlprefix\url{https://link.aps.org/doi/10.1103/PhysRevLett.58.1467}.

\bibitem[{\citenamefont{Barzykin and Gor'kov}(1995)}]{EarlyTheory1}
\bibinfo{author}{\bibfnamefont{V.}~\bibnamefont{Barzykin}} \bibnamefont{and}
  \bibinfo{author}{\bibfnamefont{L.~P.} \bibnamefont{Gor'kov}},
  \bibinfo{journal}{Phys. Rev. Lett.} \textbf{\bibinfo{volume}{74}},
  \bibinfo{pages}{4301} (\bibinfo{year}{1995}),
  \urlprefix\url{https://link.aps.org/doi/10.1103/PhysRevLett.74.4301}.

\bibitem[{\citenamefont{Ikeda and Ohashi}(1998)}]{EarlyTheory2}
\bibinfo{author}{\bibfnamefont{H.}~\bibnamefont{Ikeda}} \bibnamefont{and}
  \bibinfo{author}{\bibfnamefont{Y.}~\bibnamefont{Ohashi}},
  \bibinfo{journal}{Phys. Rev. Lett.} \textbf{\bibinfo{volume}{81}},
  \bibinfo{pages}{3723} (\bibinfo{year}{1998}),
  \urlprefix\url{https://link.aps.org/doi/10.1103/PhysRevLett.81.3723}.

\bibitem[{\citenamefont{Shah et~al.}(2000)\citenamefont{Shah, Chandra, Coleman,
  and Mydosh}}]{EarlyTheory3}
\bibinfo{author}{\bibfnamefont{N.}~\bibnamefont{Shah}},
  \bibinfo{author}{\bibfnamefont{P.}~\bibnamefont{Chandra}},
  \bibinfo{author}{\bibfnamefont{P.}~\bibnamefont{Coleman}}, \bibnamefont{and}
  \bibinfo{author}{\bibfnamefont{J.~A.} \bibnamefont{Mydosh}},
  \bibinfo{journal}{Phys. Rev. B} \textbf{\bibinfo{volume}{61}},
  \bibinfo{pages}{564} (\bibinfo{year}{2000}),
  \urlprefix\url{https://link.aps.org/doi/10.1103/PhysRevB.61.564}.

\bibitem[{\citenamefont{Chandra et~al.}(2002)\citenamefont{Chandra, Coleman,
  Mydosh, and Tripathi}}]{EarlyTheory4}
\bibinfo{author}{\bibfnamefont{P.}~\bibnamefont{Chandra}},
  \bibinfo{author}{\bibfnamefont{P.}~\bibnamefont{Coleman}},
  \bibinfo{author}{\bibfnamefont{J.~A.} \bibnamefont{Mydosh}},
  \bibnamefont{and} \bibinfo{author}{\bibfnamefont{V.}~\bibnamefont{Tripathi}},
  \bibinfo{journal}{Nature} \textbf{\bibinfo{volume}{417}},
  \bibinfo{pages}{831} (\bibinfo{year}{2002}),
  \urlprefix\url{https://doi.org/10.1038/nature00795}.

\bibitem[{\citenamefont{Mydosh}(2014)}]{Review2014}
\bibinfo{author}{\bibfnamefont{J.}~\bibnamefont{Mydosh}},
  \bibinfo{journal}{Philosophical Magazine} \textbf{\bibinfo{volume}{94}},
  \bibinfo{pages}{3640} (\bibinfo{year}{2014}).

\bibitem[{\citenamefont{Mydosh et~al.}(2020)\citenamefont{Mydosh, Oppeneer, and
  Riseborough}}]{Review2020}
\bibinfo{author}{\bibfnamefont{J.~A.} \bibnamefont{Mydosh}},
  \bibinfo{author}{\bibfnamefont{P.~M.} \bibnamefont{Oppeneer}},
  \bibnamefont{and} \bibinfo{author}{\bibfnamefont{P.~S.}
  \bibnamefont{Riseborough}}, \bibinfo{journal}{Journal of Physics: Condensed
  Matter} \textbf{\bibinfo{volume}{32}}, \bibinfo{pages}{143002}
  (\bibinfo{year}{2020}),
  \urlprefix\url{https://doi.org/10.1088/1361-648x/ab5eba}.

\bibitem[{\citenamefont{Haule and Kotliar}(2009)}]{KH2009}
\bibinfo{author}{\bibfnamefont{K.}~\bibnamefont{Haule}} \bibnamefont{and}
  \bibinfo{author}{\bibfnamefont{G.}~\bibnamefont{Kotliar}},
  \bibinfo{journal}{Nature Physics} \textbf{\bibinfo{volume}{5}},
  \bibinfo{pages}{796 EP } (\bibinfo{year}{2009}).

\bibitem[{\citenamefont{Kotetes and Varelogiannis}(2010)}]{Recent1}
\bibinfo{author}{\bibfnamefont{P.}~\bibnamefont{Kotetes}} \bibnamefont{and}
  \bibinfo{author}{\bibfnamefont{G.}~\bibnamefont{Varelogiannis}},
  \bibinfo{journal}{Phys. Rev. Lett.} \textbf{\bibinfo{volume}{104}},
  \bibinfo{pages}{106404} (\bibinfo{year}{2010}),
  \urlprefix\url{https://link.aps.org/doi/10.1103/PhysRevLett.104.106404}.

\bibitem[{\citenamefont{Dubi and Balatsky}(2011)}]{Recent2}
\bibinfo{author}{\bibfnamefont{Y.}~\bibnamefont{Dubi}} \bibnamefont{and}
  \bibinfo{author}{\bibfnamefont{A.~V.} \bibnamefont{Balatsky}},
  \bibinfo{journal}{Phys. Rev. Lett.} \textbf{\bibinfo{volume}{106}},
  \bibinfo{pages}{086401} (\bibinfo{year}{2011}),
  \urlprefix\url{https://link.aps.org/doi/10.1103/PhysRevLett.106.086401}.

\bibitem[{\citenamefont{P\'epin et~al.}(2011)\citenamefont{P\'epin, Norman,
  Burdin, and Ferraz}}]{Recent3}
\bibinfo{author}{\bibfnamefont{C.}~\bibnamefont{P\'epin}},
  \bibinfo{author}{\bibfnamefont{M.~R.} \bibnamefont{Norman}},
  \bibinfo{author}{\bibfnamefont{S.}~\bibnamefont{Burdin}}, \bibnamefont{and}
  \bibinfo{author}{\bibfnamefont{A.}~\bibnamefont{Ferraz}},
  \bibinfo{journal}{Phys. Rev. Lett.} \textbf{\bibinfo{volume}{106}},
  \bibinfo{pages}{106601} (\bibinfo{year}{2011}),
  \urlprefix\url{https://link.aps.org/doi/10.1103/PhysRevLett.106.106601}.

\bibitem[{\citenamefont{Fujimoto}(2011)}]{Recent4}
\bibinfo{author}{\bibfnamefont{S.}~\bibnamefont{Fujimoto}},
  \bibinfo{journal}{Phys. Rev. Lett.} \textbf{\bibinfo{volume}{106}},
  \bibinfo{pages}{196407} (\bibinfo{year}{2011}),
  \urlprefix\url{https://link.aps.org/doi/10.1103/PhysRevLett.106.196407}.

\bibitem[{\citenamefont{Riseborough et~al.}(2012)\citenamefont{Riseborough,
  Coqblin, and Magalh\~aes}}]{Recent5}
\bibinfo{author}{\bibfnamefont{P.~S.} \bibnamefont{Riseborough}},
  \bibinfo{author}{\bibfnamefont{B.}~\bibnamefont{Coqblin}}, \bibnamefont{and}
  \bibinfo{author}{\bibfnamefont{S.~G.} \bibnamefont{Magalh\~aes}},
  \bibinfo{journal}{Phys. Rev. B} \textbf{\bibinfo{volume}{85}},
  \bibinfo{pages}{165116} (\bibinfo{year}{2012}),
  \urlprefix\url{https://link.aps.org/doi/10.1103/PhysRevB.85.165116}.

\bibitem[{\citenamefont{Hsu and Chakravarty}(2013)}]{Recent6}
\bibinfo{author}{\bibfnamefont{C.-H.} \bibnamefont{Hsu}} \bibnamefont{and}
  \bibinfo{author}{\bibfnamefont{S.}~\bibnamefont{Chakravarty}},
  \bibinfo{journal}{Phys. Rev. B} \textbf{\bibinfo{volume}{87}},
  \bibinfo{pages}{085114} (\bibinfo{year}{2013}),
  \urlprefix\url{https://link.aps.org/doi/10.1103/PhysRevB.87.085114}.

\bibitem[{\citenamefont{Chandra et~al.}(2013)\citenamefont{Chandra, Coleman,
  and Flint}}]{Hastatic2013}
\bibinfo{author}{\bibfnamefont{P.}~\bibnamefont{Chandra}},
  \bibinfo{author}{\bibfnamefont{P.}~\bibnamefont{Coleman}}, \bibnamefont{and}
  \bibinfo{author}{\bibfnamefont{R.}~\bibnamefont{Flint}},
  \bibinfo{journal}{Nature} \textbf{\bibinfo{volume}{493}},
  \bibinfo{pages}{621} (\bibinfo{year}{2013}).

\bibitem[{\citenamefont{Jeffries et~al.}(2010)\citenamefont{Jeffries, Moore,
  Butch, and Maple}}]{Maple2010}
\bibinfo{author}{\bibfnamefont{J.~R.} \bibnamefont{Jeffries}},
  \bibinfo{author}{\bibfnamefont{K.~T.} \bibnamefont{Moore}},
  \bibinfo{author}{\bibfnamefont{N.~P.} \bibnamefont{Butch}}, \bibnamefont{and}
  \bibinfo{author}{\bibfnamefont{M.~B.} \bibnamefont{Maple}},
  \bibinfo{journal}{Phys. Rev. B} \textbf{\bibinfo{volume}{82}},
  \bibinfo{pages}{033103} (\bibinfo{year}{2010}),
  \urlprefix\url{https://link.aps.org/doi/10.1103/PhysRevB.82.033103}.

\bibitem[{\citenamefont{Park et~al.}(2012)\citenamefont{Park, Tobash, Ronning,
  Bauer, Sarrao, Thompson, and Greene}}]{Fano2012}
\bibinfo{author}{\bibfnamefont{W.~K.} \bibnamefont{Park}},
  \bibinfo{author}{\bibfnamefont{P.~H.} \bibnamefont{Tobash}},
  \bibinfo{author}{\bibfnamefont{F.}~\bibnamefont{Ronning}},
  \bibinfo{author}{\bibfnamefont{E.~D.} \bibnamefont{Bauer}},
  \bibinfo{author}{\bibfnamefont{J.~L.} \bibnamefont{Sarrao}},
  \bibinfo{author}{\bibfnamefont{J.~D.} \bibnamefont{Thompson}},
  \bibnamefont{and} \bibinfo{author}{\bibfnamefont{L.~H.}
  \bibnamefont{Greene}}, \bibinfo{journal}{Phys. Rev. Lett.}
  \textbf{\bibinfo{volume}{108}}, \bibinfo{pages}{246403}
  (\bibinfo{year}{2012}),
  \urlprefix\url{https://link.aps.org/doi/10.1103/PhysRevLett.108.246403}.

\bibitem[{\citenamefont{Lu et~al.}(2012)\citenamefont{Lu, Ronning, Tobash,
  Gofryk, Bauer, and Thompson}}]{Lu2012}
\bibinfo{author}{\bibfnamefont{X.}~\bibnamefont{Lu}},
  \bibinfo{author}{\bibfnamefont{F.}~\bibnamefont{Ronning}},
  \bibinfo{author}{\bibfnamefont{P.~H.} \bibnamefont{Tobash}},
  \bibinfo{author}{\bibfnamefont{K.}~\bibnamefont{Gofryk}},
  \bibinfo{author}{\bibfnamefont{E.~D.} \bibnamefont{Bauer}}, \bibnamefont{and}
  \bibinfo{author}{\bibfnamefont{J.~D.} \bibnamefont{Thompson}},
  \bibinfo{journal}{Phys. Rev. B} \textbf{\bibinfo{volume}{85}},
  \bibinfo{pages}{020402} (\bibinfo{year}{2012}),
  \urlprefix\url{https://link.aps.org/doi/10.1103/PhysRevB.85.020402}.

\bibitem[{\citenamefont{Kung et~al.}(2015)\citenamefont{Kung, Baumbach, Bauer,
  Thorsmølle, Zhang, Haule, Mydosh, and Blumberg}}]{Girsh2015}
\bibinfo{author}{\bibfnamefont{H.-H.} \bibnamefont{Kung}},
  \bibinfo{author}{\bibfnamefont{R.~E.} \bibnamefont{Baumbach}},
  \bibinfo{author}{\bibfnamefont{E.~D.} \bibnamefont{Bauer}},
  \bibinfo{author}{\bibfnamefont{V.~K.} \bibnamefont{Thorsmølle}},
  \bibinfo{author}{\bibfnamefont{W.-L.} \bibnamefont{Zhang}},
  \bibinfo{author}{\bibfnamefont{K.}~\bibnamefont{Haule}},
  \bibinfo{author}{\bibfnamefont{J.~A.} \bibnamefont{Mydosh}},
  \bibnamefont{and} \bibinfo{author}{\bibfnamefont{G.}~\bibnamefont{Blumberg}},
  \bibinfo{journal}{Science} \textbf{\bibinfo{volume}{347}},
  \bibinfo{pages}{1339} (\bibinfo{year}{2015}),
  \eprint{https://www.science.org/doi/pdf/10.1126/science.1259729},
  \urlprefix\url{https://www.science.org/doi/abs/10.1126/science.1259729}.

\bibitem[{\citenamefont{Wolowiec et~al.}(2021)\citenamefont{Wolowiec,
  Kanchanavatee, Huang, Ran, Breindel, Pouse, Sasmal, Baumbach, Chappell,
  Riseborough et~al.}}]{PNAS1}
\bibinfo{author}{\bibfnamefont{C.~T.} \bibnamefont{Wolowiec}},
  \bibinfo{author}{\bibfnamefont{N.}~\bibnamefont{Kanchanavatee}},
  \bibinfo{author}{\bibfnamefont{K.}~\bibnamefont{Huang}},
  \bibinfo{author}{\bibfnamefont{S.}~\bibnamefont{Ran}},
  \bibinfo{author}{\bibfnamefont{A.~J.} \bibnamefont{Breindel}},
  \bibinfo{author}{\bibfnamefont{N.}~\bibnamefont{Pouse}},
  \bibinfo{author}{\bibfnamefont{K.}~\bibnamefont{Sasmal}},
  \bibinfo{author}{\bibfnamefont{R.~E.} \bibnamefont{Baumbach}},
  \bibinfo{author}{\bibfnamefont{G.}~\bibnamefont{Chappell}},
  \bibinfo{author}{\bibfnamefont{P.~S.} \bibnamefont{Riseborough}},
  \bibnamefont{et~al.}, \bibinfo{journal}{Proceedings of the National Academy
  of Sciences} \textbf{\bibinfo{volume}{118}} (\bibinfo{year}{2021}), ISSN
  \bibinfo{issn}{0027-8424}.

\bibitem[{\citenamefont{Ran et~al.}(2016)\citenamefont{Ran, Wolowiec, Jeon,
  Pouse, Kanchanavatee, White, Huang, Martien, DaPron, Snow et~al.}}]{PNAS2}
\bibinfo{author}{\bibfnamefont{S.}~\bibnamefont{Ran}},
  \bibinfo{author}{\bibfnamefont{C.~T.} \bibnamefont{Wolowiec}},
  \bibinfo{author}{\bibfnamefont{I.}~\bibnamefont{Jeon}},
  \bibinfo{author}{\bibfnamefont{N.}~\bibnamefont{Pouse}},
  \bibinfo{author}{\bibfnamefont{N.}~\bibnamefont{Kanchanavatee}},
  \bibinfo{author}{\bibfnamefont{B.~D.} \bibnamefont{White}},
  \bibinfo{author}{\bibfnamefont{K.}~\bibnamefont{Huang}},
  \bibinfo{author}{\bibfnamefont{D.}~\bibnamefont{Martien}},
  \bibinfo{author}{\bibfnamefont{T.}~\bibnamefont{DaPron}},
  \bibinfo{author}{\bibfnamefont{D.}~\bibnamefont{Snow}}, \bibnamefont{et~al.},
  \bibinfo{journal}{Proceedings of the National Academy of Sciences}
  \textbf{\bibinfo{volume}{113}}, \bibinfo{pages}{13348}
  (\bibinfo{year}{2016}).

\bibitem[{\citenamefont{Ran et~al.}(2017)\citenamefont{Ran, Jeon, Pouse,
  Breindel, Kanchanavatee, Huang, Gallagher, Chen, Graf, Baumbach
  et~al.}}]{PNAS3}
\bibinfo{author}{\bibfnamefont{S.}~\bibnamefont{Ran}},
  \bibinfo{author}{\bibfnamefont{I.}~\bibnamefont{Jeon}},
  \bibinfo{author}{\bibfnamefont{N.}~\bibnamefont{Pouse}},
  \bibinfo{author}{\bibfnamefont{A.~J.} \bibnamefont{Breindel}},
  \bibinfo{author}{\bibfnamefont{N.}~\bibnamefont{Kanchanavatee}},
  \bibinfo{author}{\bibfnamefont{K.}~\bibnamefont{Huang}},
  \bibinfo{author}{\bibfnamefont{A.}~\bibnamefont{Gallagher}},
  \bibinfo{author}{\bibfnamefont{K.-W.} \bibnamefont{Chen}},
  \bibinfo{author}{\bibfnamefont{D.}~\bibnamefont{Graf}},
  \bibinfo{author}{\bibfnamefont{R.~E.} \bibnamefont{Baumbach}},
  \bibnamefont{et~al.}, \bibinfo{journal}{Proceedings of the National Academy
  of Sciences} \textbf{\bibinfo{volume}{114}}, \bibinfo{pages}{9826}
  (\bibinfo{year}{2017}), ISSN \bibinfo{issn}{0027-8424}.

\bibitem[{\citenamefont{Haule and Kotliar}(2010)}]{KH2010}
\bibinfo{author}{\bibfnamefont{K.}~\bibnamefont{Haule}} \bibnamefont{and}
  \bibinfo{author}{\bibfnamefont{G.}~\bibnamefont{Kotliar}},
  \bibinfo{journal}{EPL (Europhysics Letters)} \textbf{\bibinfo{volume}{89}},
  \bibinfo{pages}{57006} (\bibinfo{year}{2010}).

\bibitem[{\citenamefont{Bardeen et~al.}(1957)\citenamefont{Bardeen, Cooper, and
  Schrieffer}}]{BCS}
\bibinfo{author}{\bibfnamefont{J.}~\bibnamefont{Bardeen}},
  \bibinfo{author}{\bibfnamefont{L.~N.} \bibnamefont{Cooper}},
  \bibnamefont{and} \bibinfo{author}{\bibfnamefont{J.~R.}
  \bibnamefont{Schrieffer}}, \bibinfo{journal}{Phys. Rev.}
  \textbf{\bibinfo{volume}{108}}, \bibinfo{pages}{1175} (\bibinfo{year}{1957}).

\bibitem[{\citenamefont{Eliashberg}(1986)}]{Eliashberg1986}
\bibinfo{author}{\bibfnamefont{G.~M.} \bibnamefont{Eliashberg}},
  \bibinfo{journal}{Pis'ma Zh. Eksp. Teor. Phys.}
  \textbf{\bibinfo{volume}{45}}, \bibinfo{pages}{28} (\bibinfo{year}{1986}).

\bibitem[{\citenamefont{Kos et~al.}(2004)\citenamefont{Kos, Millis, and
  Larkin}}]{Kos2004}
\bibinfo{author}{\bibfnamefont{S.}~\bibnamefont{Kos}},
  \bibinfo{author}{\bibfnamefont{A.~J.} \bibnamefont{Millis}},
  \bibnamefont{and} \bibinfo{author}{\bibfnamefont{A.~I.}
  \bibnamefont{Larkin}}, \bibinfo{journal}{Phys. Rev. B}
  \textbf{\bibinfo{volume}{70}}, \bibinfo{pages}{214531}
  (\bibinfo{year}{2004}).

\bibitem[{\citenamefont{Fischer et~al.}(2018)\citenamefont{Fischer, Hecker,
  Hoyer, and Schmalian}}]{Joerg1}
\bibinfo{author}{\bibfnamefont{S.}~\bibnamefont{Fischer}},
  \bibinfo{author}{\bibfnamefont{M.}~\bibnamefont{Hecker}},
  \bibinfo{author}{\bibfnamefont{M.}~\bibnamefont{Hoyer}}, \bibnamefont{and}
  \bibinfo{author}{\bibfnamefont{J.}~\bibnamefont{Schmalian}},
  \bibinfo{journal}{Phys. Rev. B} \textbf{\bibinfo{volume}{97}},
  \bibinfo{pages}{054510} (\bibinfo{year}{2018}),
  \urlprefix\url{https://link.aps.org/doi/10.1103/PhysRevB.97.054510}.

\bibitem[{\citenamefont{Hoyer and Schmalian}(2018)}]{Joerg2}
\bibinfo{author}{\bibfnamefont{M.}~\bibnamefont{Hoyer}} \bibnamefont{and}
  \bibinfo{author}{\bibfnamefont{J.}~\bibnamefont{Schmalian}},
  \bibinfo{journal}{Phys. Rev. B} \textbf{\bibinfo{volume}{97}},
  \bibinfo{pages}{224423} (\bibinfo{year}{2018}),
  \urlprefix\url{https://link.aps.org/doi/10.1103/PhysRevB.97.224423}.

\bibitem[{\citenamefont{Shen and Dzero}(2018)}]{Shen2018}
\bibinfo{author}{\bibfnamefont{P.}~\bibnamefont{Shen}} \bibnamefont{and}
  \bibinfo{author}{\bibfnamefont{M.}~\bibnamefont{Dzero}},
  \bibinfo{journal}{Phys. Rev. B} \textbf{\bibinfo{volume}{98}},
  \bibinfo{pages}{125131} (\bibinfo{year}{2018}),
  \urlprefix\url{https://link.aps.org/doi/10.1103/PhysRevB.98.125131}.

\bibitem[{\citenamefont{Vaks et~al.}(1967)\citenamefont{Vaks, Larkin, and
  Pikin}}]{VLP}
\bibinfo{author}{\bibfnamefont{V.~G.} \bibnamefont{Vaks}},
  \bibinfo{author}{\bibfnamefont{A.~I.} \bibnamefont{Larkin}},
  \bibnamefont{and} \bibinfo{author}{\bibfnamefont{S.~A.} \bibnamefont{Pikin}},
  \bibinfo{journal}{Sov. Phys. - JETP} \textbf{\bibinfo{volume}{24}},
  \bibinfo{pages}{240} (\bibinfo{year}{1967}).

\bibitem[{\citenamefont{Das et~al.}(2015)\citenamefont{Das, Kanchanavatee,
  Helton, Huang, Baumbach, Bauer, White, Burnett, Maple, Lynn
  et~al.}}]{Pinaki2015}
\bibinfo{author}{\bibfnamefont{P.}~\bibnamefont{Das}},
  \bibinfo{author}{\bibfnamefont{N.}~\bibnamefont{Kanchanavatee}},
  \bibinfo{author}{\bibfnamefont{J.~S.} \bibnamefont{Helton}},
  \bibinfo{author}{\bibfnamefont{K.}~\bibnamefont{Huang}},
  \bibinfo{author}{\bibfnamefont{R.~E.} \bibnamefont{Baumbach}},
  \bibinfo{author}{\bibfnamefont{E.~D.} \bibnamefont{Bauer}},
  \bibinfo{author}{\bibfnamefont{B.~D.} \bibnamefont{White}},
  \bibinfo{author}{\bibfnamefont{V.~W.} \bibnamefont{Burnett}},
  \bibinfo{author}{\bibfnamefont{M.~B.} \bibnamefont{Maple}},
  \bibinfo{author}{\bibfnamefont{J.~W.} \bibnamefont{Lynn}},
  \bibnamefont{et~al.}, \bibinfo{journal}{Phys. Rev. B}
  \textbf{\bibinfo{volume}{91}}, \bibinfo{pages}{085122}
  (\bibinfo{year}{2015}),
  \urlprefix\url{https://link.aps.org/doi/10.1103/PhysRevB.91.085122}.

\end{thebibliography}

\end{document}